\documentclass[
aps,
prl,
twocolumn,
superscriptaddress,
longbibliography
]{revtex4-2}

\usepackage{graphicx}
\usepackage{amsmath,amssymb,bm}
\usepackage{physics}
\usepackage[colorlinks=true,allcolors=blue]{hyperref}

\begin{document}

\title{Hidden universality in dislocation-loops--mediated three-dimensional crystal melting}

\author{Alessio Zaccone}
\email{alessio.zaccone@unimi.it}
\affiliation{Department of Physics ``A.~Pontremoli'', University of Milan, via Celoria 16, 20133 Milan, Italy}

\author{Konrad Samwer}
\email{ksamwer@gwdg.de}
\affiliation{I.~Physikalisches Institut, University of G\"ottingen, Friedrich-Hund-Platz 1, 37077 G\"ottingen, Germany}

\date{\today}

\begin{abstract}
Understanding why and how crystalline solids melt remains a central problem in condensed-matter physics.
Dislocation loops are fundamental topological excitations that control the thermodynamic stability of crystals, yet their role in setting universal aspects of melting has remained unclear.
Here we show, within dislocation-mediated melting theory, that the free-energy condition for loop proliferation leads to a universal ratio between the energy of a minimal dislocation loop and the thermal energy at melting.
For minimal dislocation loops that begin to proliferate at the onset of melting, this ratio takes the purely geometric value
$\mathcal{E}_* = E_{\rm loop}/(k_B T_m) \approx 25.1$, independent of elastic moduli and chemistry-dependent details.
This result provides a microscopic explanation for recent empirical findings by Lunkenheimer \emph{et al.}, who identified a closely related universal energy scale $\approx 24.6$ from viscosity data.
The same framework also rationalizes the empirical $2/3$ rule relating the glass-transition and melting temperatures.
\end{abstract}
\maketitle

The microscopic mechanism by which crystalline solids melt remains a long-standing challenge in condensed-matter physics~\cite{deWith}.
Early theoretical approaches described melting as a mechanical instability of the lattice, associated with phonon softening or the collapse of elastic constants near the melting temperature~\cite{Born_melting,Born_Huang,Ma,yip,Zaccone_book}.
While successful in specific contexts, such quasi-harmonic descriptions fail to capture the essential role of topological defects, large-amplitude collective rearrangements~\cite{Rongchao}, and disorder~\cite{Lunkenheimer2023}, which accompany melting in real materials~\cite{Zippelius}.

A more fundamental perspective interprets melting as a defect-unbinding transition driven by the proliferation of dislocations.
This viewpoint was developed in a unified form by Kleinert through his gauge-field theory of stresses and defects in solids~\cite{Kleinert1989}, in which dislocations emerge as dynamical degrees of freedom and melting occurs when the configurational entropy of defect loops compensates their elastic and core energies.
Subsequent analytical work by Burakovsky, Preston and collaborators demonstrated that dislocation-mediated models can quantitatively reproduce melting temperatures and latent heats of crystalline metals with different lattice symmetries~\cite{BurakovskyPreston1999,BurakovskyPreston2000}, building on classical dislocation elasticity~\cite{GranatoLucke1956,Nabarro1967}.

The notion of melting as a defect-proliferation transition echoes the unbinding of vortices in two-dimensional superfluids~\cite{KosterlitzThouless1973} and the Halperin--Nelson theory of two-dimensional crystal melting~\cite{HalperinNelson1978}.
In three dimensions, however, the relevant excitations are closed dislocation loops, whose energetics and entropy jointly determine the stability of the crystalline phase.
Despite decades of work, the extent to which this picture implies universal features of three-dimensional melting has remained largely unexplored.

Molecular-dynamics simulations provide strong support for a dislocation-mediated scenario, directly revealing the nucleation and proliferation of dislocation segments and loops as melting is approached~\cite{Phillpot1991,HoytAstaKarma2001,Zhakhovskii1999,Kadau2002,Bodapati2006}.
These studies show that melting is preceded by intense defect activity, shear localisation, and the emergence of configurational excitations~\cite{Ojovan,Egami,Johnson}, consistent with a topological interpretation of the transition.

We demonstrate that three-dimensional crystal melting is governed by a previously unrecognized universal energy scale.
Specifically, we show that at the melting temperature the ratio between the energy of a minimal dislocation loop and the thermal energy $k_B T_m$ becomes a geometry-controlled constant, $\mathcal{E}_* \approx 25.1$.
Remarkably, this ratio is independent of elastic constants, core energies, and chemical details.
We further show that this universal constant naturally connects to recent work by Lunkenheimer, Samwer and Loidl~\cite{Lunkenheimer2025}, who identified a universal cooperativity-free activation energy at melting corresponding to a closely related value $\approx 24.6$.

\begin{figure*}[t]
\centering
\includegraphics[width=0.9\textwidth]{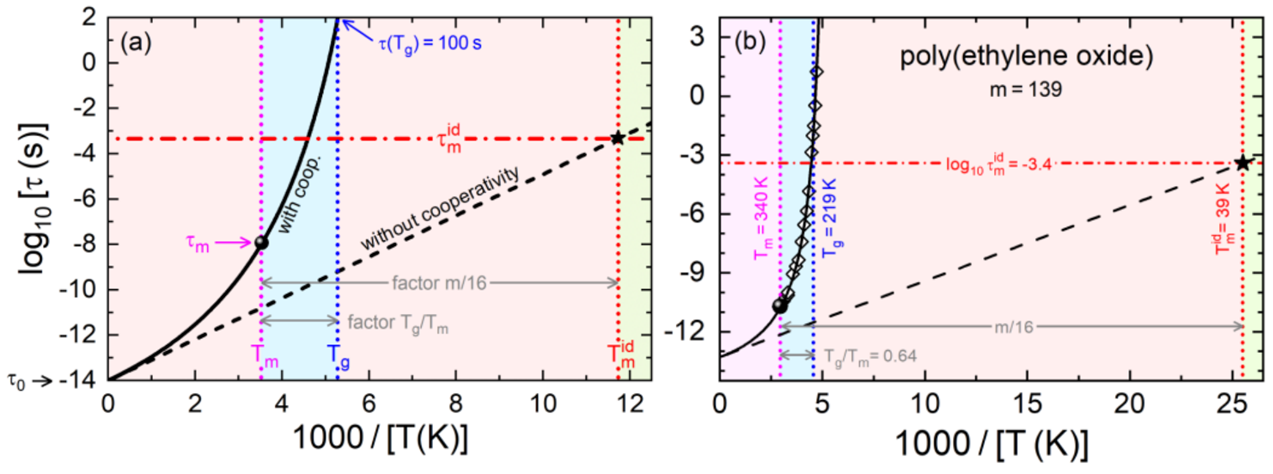}
\caption{Construction of the idealized, cooperativity-free melting point and the corresponding universal relaxation time.
(a) Schematic Arrhenius plot of the relaxation time $\tau(T)$. The solid line represents the experimentally observed non-Arrhenius (VFT) behavior including cooperativity, while the dashed line shows the extrapolated Arrhenius behavior expected in the absence of cooperativity.
(b) Same construction applied to poly(ethylene oxide) ($m=139$).
Reproduced with permission from Ref.~\cite{Lunkenheimer2025}. 
Copyright (2025) by the American Physical Society.}
\label{fig1}
\end{figure*}

The universal energy ratio $24.6$ at 3D melting follows directly from the Arrhenius construction shown in Fig. \ref{fig1} and does not involve any adjustable parameter. In Fig. \ref{fig1}(a), besides the non-Arrhenius Vogel-Fulcher-Tammann (VFT) curve, a hypothetical Arrhenius line is drawn to represent the dynamics of the liquid in the absence of cooperativity. This line is written as
\begin{equation}
y(T)=y_0 \exp\!\left(\frac{E}{T}\right),
\end{equation}
where $E$ is an effective activation barrier expressed in Kelvin. Here $y(T)$ denotes a generic dynamical quantity characterizing molecular mobility in the liquid. Depending on context, $y$ represents either the structural relaxation time $\tau(T)$ or the viscosity $\eta(T)$. Both quantities are treated on equal footing because they exhibit very similar temperature dependencies over many decades and are approximately proportional to each other in glass-forming liquids.

The idealized melting temperature $T_m^{\mathrm{id}}$ is defined as the temperature at which this Arrhenius line reaches the universal relaxation time identified earlier in Ref. \cite{Lunkenheimer2025},
\begin{equation}
\log_{10}\tau_m^{\mathrm{id}} = -3.33,
\end{equation}
indicated by the star in Fig. \ref{fig1}(a). Evaluating the Arrhenius law at $T=T_m^{\mathrm{id}}$ yields
\begin{equation}
\log_{10} y_m^{\mathrm{id}} - \log_{10} y_0
= \frac{E}{T_m^{\mathrm{id}} \ln 10}.
\end{equation}

Using the standard assumptions employed throughout the analysis, $\log_{10} y_0=-14$ for relaxation times, one obtains
\begin{equation}
\log_{10} y_m^{\mathrm{id}} - \log_{10} y_0
= -3.33 - (-14) = 10.67.
\end{equation}
Multiplication by $\ln 10$ then gives
\begin{equation}
\frac{E}{T_m^{\mathrm{id}}} = 10.67\,\ln 10 \approx 24.6,
\end{equation}
which leads directly to,
\begin{equation}
E = 24.6\, T_m^{\mathrm{id}}.\label{ratio}
\end{equation}

Thus, the factor $24.6$ is simply the dimensionless slope of the Arrhenius line in Fig.~1, corresponding to the logarithmic separation between the microscopic time scale $y_0$ and the universal ideal melting point. For illustration purposes, this construction is shown in Fig. \ref{fig1}(b) for a specific material, polyethylene oxide. In Ref. \cite{Lunkenheimer2025}, the authors have shown that the same construction applied to a variety of different materials always yields the same result, and the universal, constant energy ratio of 24.6 in Eq. \eqref{ratio}.

In what follows, we shall present a mathematical derivation of how this constant energy ratio arises from a dislocation-loop theory of 3D melting.

Within dislocation-mediated melting theory, melting occurs when the free energy
of a dislocation loop vanishes due to the compensation between elastic energy
and configurational entropy.
As shown in the Appendix, for a circular loop of radius $R$ the free-energy
condition $F(R,T_m)=0$ leads to a melting temperature
$k_B T_m = a G b^2 D(R)/\ln z$, where $D(R)$ contains elastic and core
contributions.
Importantly, all elastic constants and core parameters cancel from the
dimensionless ratio $E_{\rm loop}/(k_B T_m)$, leaving a purely geometric result.

We therefore obtain the central result
\begin{equation}
\frac{E_{\rm loop}}{k_B T_m}
= \frac{2\pi R}{a}\,\ln z,
\label{eq:El_over_kBTm}
\end{equation}
which is independent of elastic moduli, core energies, and chemical details,
and depends only on geometry and local coordination.

For the smallest mechanically stable dislocation loops,
atomistic simulations give $R_{\min}\sim 1\,\mathrm{nm}$, while
$a\sim 0.3\,\mathrm{nm}$.
Using $\ln z\simeq 1.2$ for typical local coordination, Eq.~\eqref{eq:El_over_kBTm}
yields
\begin{equation}
\mathcal{E}_* \equiv \frac{E_{\rm loop}}{k_B T_m} \approx 25.1.
\end{equation}
This universal value is independent of elastic moduli and microscopic
interaction details.

\begin{figure}[t]
\centering
\includegraphics[width=\linewidth]{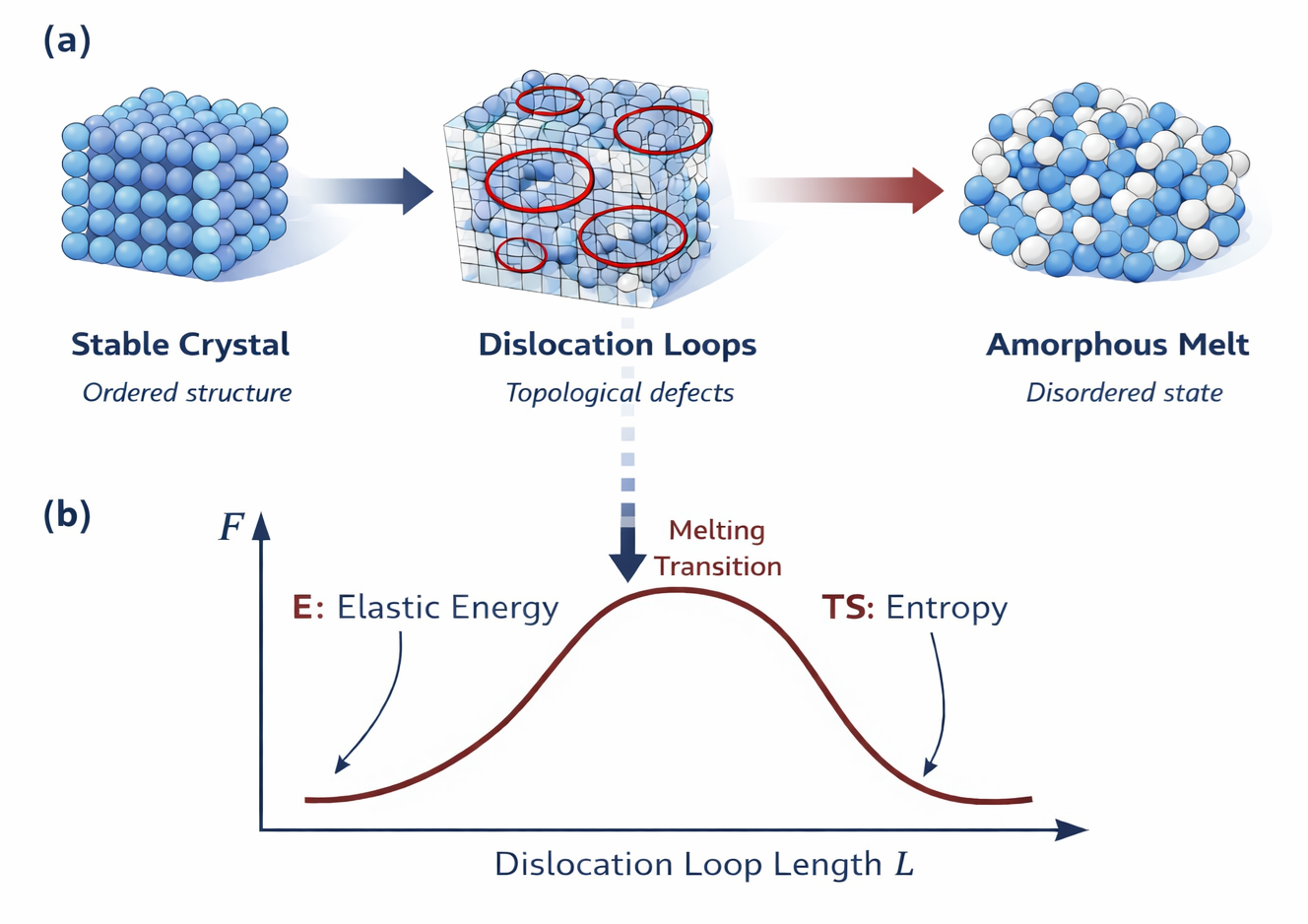}
\caption{Mechanism of dislocation-loop--mediated melting in three-dimensional crystals.}
\label{fig2}
\end{figure}


Atomistic calculations and molecular-dynamics simulations consistently show that the smallest mechanically stable dislocation loops have radii of order 1 nm, with smaller loops collapsing due to line tension and core-energy effects \cite{Phillpot1991,Zhakhovskii1999,Bodapati2006}. Taken together, these studies justify adopting $R_{\min} \sim 1\,\mathrm{nm}$ as the smallest physically meaningful loop radius entering the dislocation-mediated melting criterion. Interestingly, the value of about 2-3 nm (i.e. twice $R_{\textrm{min}}$) is also largely compatible with the universal size of cooperative rearranging regions in supercooled liquids undergoing glass transition that has been recently found in Ref. \cite{Loeffler}.

At the \emph{idealized} melting the energy ratio must be just the same as the one derived above, because at the idealized melting there is no effect of cooperativity nor of any other chemistry-dependent factor. Therefore, we have:
\begin{equation}
\mathcal{E}_* \equiv \frac{E_{\rm loop}}{k_B T_m} \equiv \frac{E}{k_B T_m^{\mathrm{id}}},
\label{eq:Edim_id}
\end{equation}
where $\frac{E}{k_B T_m^{\mathrm{id}}}$ is the energy ratio analyzed and discussed in Ref. \cite{Lunkenheimer2025}.

While the dimensionless constant Eq.~\eqref{eq:Edim_id} is universal, the melting temperature itself retains its dependence on materials parameters. From Eq.~\eqref{eq:Tm_general} in the Appendix, evaluated at $R=R_{\min}$, we can write
\begin{equation}
T_m
= \frac{
G b^{2} a
}{
2\pi k_B \ln z
}
\left[
C_1 \ln\!\left(\frac{R_{\min}}{r_c}\right)
+ C_2
\right],
\label{eq:Tm_final}
\end{equation}
which explicitly involves $G$, $\nu$ (through $C_1$), $b$, and details of the core structure (through $C_2$). Equivalently, using Eq.~\eqref{eq:kBTm_explicit}, derived in the Appendix and evaluated at $R=R_{\min}$,
\begin{equation}
k_B T_m
= a G b^{2}
\frac{
\displaystyle \frac{1}{2(1-\nu)}
\ln\!\left(\frac{R_{\min}}{r_c}\right) + \alpha
}{
\ln z
}.
\label{eq:kBTm_pdf}
\end{equation}
This formula cannot be expected to be quantitatively accurate, but it provides a useful description of the chemistry-dependent parameters, which qualitatively control the melting phenomenon.

Although the dislocation core-energy parameter $\alpha$ enters the absolute value of the melting temperature, through $C_2$, it cancels exactly from the dimensionless ratio $E_{\rm loop}/(k_B T_m)$. The universal constant that we shall derive below therefore does not depend on microscopic details of the dislocation core.

Lunkenheimer, Samwer, and Loidl~\cite{Lunkenheimer2025} have recently revisited the 2/3 rule from a complementary perspective. Starting from the empirical observation that $T_g/T_m \approx 2/3$ for a broad class of glass-forming liquids, they show that the wide variation (over many decades) of viscosities and relaxation times at the \emph{experimental} melting temperature $T_m$ can be traced back to differences in liquid fragility and cooperativity. By conceptually removing cooperativity, they construct an ``idealized,'' fragility-free melting point at which all crystals would melt into liquids with a \emph{universal} viscosity and relaxation time.   
Their analysis implies a universal activation barrier for viscous flow at melting, which can be expressed as a dimensionless constant $\frac{E_{\eta}}{k_B T_m} \approx 24.6$,
where $E_{\eta}$ characterizes the effective energy scale controlling the viscosity or structural relaxation at the (idealized) melting point. While the detailed definition of $E_{\eta}$ differs from the dislocation-loop energy $E_{\rm loop}$ used here, the closeness of the two dimensionless values,
\begin{equation}
\mathcal{E}_* \approx 25.1,
\qquad
\frac{E_{\eta}}{k_B T_m} \approx 24.6,
\label{eq:comparison}
\end{equation}
strongly suggests that both approaches are probing the \emph{same underlying universal energy scale} associated with the onset of cooperative, defect-mediated motion at melting.

In our framework, this universal energy arises from the entropy of dislocation-loop configurations, Eq.~\eqref{eq:Edim_R} derived in the Appendix, and is controlled only by the ratio $R_{\min}/a$ and the local coordination $z$. In the Lunkenheimer--Samwer--Loidl picture, the universal constant emerges from the suppression of fragility and cooperativity, revealing a ``bare'' melting process characterized by a single energy scale. The agreement at the 2--3\% level between $25.1$ and $24.6$ is well within the expected uncertainty due to approximate choices of $R_{\min}$, $a$, and $z$ in our estimate, and to the phenomenological nature of the analysis in Ref.~\cite{Lunkenheimer2025}.

Taken together, the two approaches---one defect-based, one dynamical---offer a coherent and mutually reinforcing picture. Indeed, the 2/3 rule (previously rationalized within the Lindemann and Born melting pictures \cite{ZacconeJCP}), the near-universality of a dimensionless melting energy scale, and the universality of viscosity in the absence of fragility \cite{Lunkenheimer2025}, all originate from the same topological mechanism of melting driven by dislocation loops.

The parameter $z$ reflects the number of local configurational choices (effectively, coordination) available to a segment of the dislocation line. In a crystalline environment, such as a simple cubic lattice, one has
\begin{equation}
z_{\mathrm{crystal}} = 6,
\qquad
\ln z_{\mathrm{crystal}} = \ln(6) = 1.791769\ldots.
\label{eq:lnz_crystal}
\end{equation}
In a supercooled dense liquid, local coordination can reach up to, at most, $13$ nearest neighbors \cite{Hansen}, giving
\begin{equation}
z_{\mathrm{glass}} = 13,
\qquad
\ln z_{\mathrm{glass}} = \ln(13) = 2.564949\ldots.
\label{eq:lnz_glass}
\end{equation}
The ratio of configurational entropies (per segment) is therefore
\begin{equation}
\frac{\ln z_{\mathrm{crystal}}}{\ln z_{\mathrm{glass}}}
= \frac{\ln 6}{\ln 13}
\approx 0.6985\ldots,
\label{eq:lnz_ratio}
\end{equation}
which is remarkably close to the empirical relation
\begin{equation}
\frac{T_g}{T_m} \approx \frac{2}{3} = 0.666\ldots.
\label{eq:2over3}
\end{equation}
This suggests that the 2/3 rule can be understood as arising from the increase in configurational entropy available to defects in the glass relative to the crystal: the higher $z$ in the disordered state lowers the temperature at which defect proliferation becomes favorable, yielding $T_g < T_m$ with a ratio set primarily by the ratio of $\ln z$ values.

In the present framework,
$\mathcal{E}_*$ does not represent an elastic energy scale but, at the melting
threshold $F=0$, is exactly equal to the configurational entropy of the
smallest proliferating dislocation loops in units of $k_B$. Melting occurs
when this entropy overcomes the energetic cost of creating topological
excitations, causing elastic constants and core parameters to cancel from the
dimensionless ratio.

Fragility-based approaches probe the same underlying physics, in supercooled liquids, from a dynamical
perspective. Within the Krausser-Samwer-Zaccone (KSZ) model \cite{KSZ}, fragility is controlled by the steepness of
the effective medium-range repulsion and governs the activation of cooperative
shear rearrangements. As shown in the Appendix, both the fragility index $m$ and
the KSZ repulsion parameter $\lambda$ increase monotonically with the Poisson
ratio $\nu$ and diverge in the incompressible limit $\nu \rightarrow 0.5$,
corresponding to infinitely steep, hard-wall repulsion. Steeper repulsion
reduces configurational freedom, thereby increasing the energetic barrier for
cooperative motion.

The universality of $\mathcal{E}_*$ reflects the fact that, irrespective of
elastic stiffness or interaction steepness, melting requires a fixed minimum
amount of configurational entropy associated with topology-changing motion.
The agreement between $\mathcal{E}_* \approx 25.1$ and
$E_{\eta}/(k_B T_m) \approx 24.6$ therefore suggests that both dislocation-loop
proliferation and viscous flow at melting are governed by the same fundamental,
entropy-controlled energy scale.

We have shown that dislocation-mediated melting in three dimensions is governed by a universal, dimensionless energy scale. By analysing the free-energy balance between the elastic cost and configurational entropy of dislocation loops, we find that at the melting temperature the ratio $E_{\rm loop}/(k_B T_m)$ equals the loop entropy in units of $k_B$, yielding the purely geometric expression
\[
\mathcal{E}(R) = \frac{2\pi R}{a}\ln z.
\]
For minimal, nanometre-scale loops this predicts the universal value $\mathcal{E}_* \approx 25.1$, independent of elastic moduli, core energies, and microscopic bonding details.

Beyond the noninteracting-loop picture adopted here, it is conceivable that long-range Eshelby-type elastic interactions between dislocation loops could give rise to collective effects such as loop clustering or effective condensation near melting, an intriguing possibility that lies beyond the scope of the present work.

This result provides a microscopic foundation for recent experimental observations by Lunkenheimer, Samwer and Loidl \cite{Lunkenheimer2025}, who identified a closely related universal energy scale $\approx 24.6$ from viscosity data at melting across a wide range of materials. The close numerical agreement between these values strongly suggests that both approaches probe the same underlying physics: the onset of defect-mediated, topology-changing motion at melting. Within this framework, the empirical $2/3$ rule relating the glass-transition and melting temperatures emerges naturally from differences in configurational entropy between crystalline and disordered states.

More broadly, our findings reveal an unexpected level of universality in three-dimensional melting. This insight unifies melting, viscous flow, and glass formation within a common topological framework and opens new perspectives on the fundamental mechanisms controlling phase stability in condensed matter.

\section*{Acknowledgements}

K.S. acknowledges discussions with P.~Lunkenheimer and A.~Loidl.
A.Z. acknowledges funding from the European Research Council (Grant No.~101043968),
the US Army Research Office (W911NF-22-2-0256),
and the Nieders\"achsische Akademie der Wissenschaften zu G\"ottingen.

\bibliographystyle{apsrev4-2}
\bibliography{refs}

\appendix

\section{End Matter}

\section{Dislocation energetics}
We follow Kleinert's continuum-elasticity treatment of dislocations, supplemented by a short-range core energy~\cite{Kleinert1989}. Consider a circular dislocation loop of radius $R$, line length
\begin{equation}
L = 2\pi R,
\label{eq:L}
\end{equation}
and Burgers vector magnitude $b$. The total energy of the loop is written as the sum of an elastic and a core contribution,
\begin{equation}
E_{\rm tot} = E_{\rm el} + E_{\rm core}.
\label{eq:Etot}
\end{equation}
The total energy $E_{\rm tot}$ contains, in principle, a long-range contribution of Eshelby type arising from elastic interactions between distant regions; however, this non-local term is neglected here since it mainly renormalizes the background elastic energy and does not affect the local mechanisms or qualitative behavior under consideration (see e.g. \cite{Eshelby1957,FalkLanger1998}). While such long-range contributions may be important in certain conditions (e.g. for shear banding \cite{Wilde}), they may also be screened as the system undergoes disordering due to defects, fluctuations etc. \cite{Moshe}.
To logarithmic accuracy, the elastic energy of a dislocation segment of length $L$ is
\begin{equation}
E_{\rm el}
\approx \frac{G b^{2}}{4\pi(1-\nu)}\,L \ln\!\left(\frac{R}{r_c}\right),
\label{eq:Eel_segment}
\end{equation}
where $G$ is the shear modulus, $\nu$ the Poisson ratio, and $r_c$ a core radius of order $b$. 
Equation~\eqref{eq:Eel_segment} is the standard logarithmic expression for the elastic energy of a dislocation line in isotropic elasticity, derived in classical dislocation theory and valid to logarithmic accuracy in the ratio of the loop radius to the core radius~\cite{HirthLothe2017,Nabarro1967,Kleinert1989}.

For a circular loop with $R=L/2\pi$, this becomes
\begin{equation}
E_{{\rm loop},{\rm el}}(R)
= \frac{G b^{2}}{2(1-\nu)}\,R \ln\!\left(\frac{R}{r_c}\right).
\label{eq:Eel_loop}
\end{equation}
The core contribution is taken to be linear in $L$ \cite{HirthLothe2017},
\begin{equation}
E_{\rm core} = \alpha\,G b^{2} L,
\label{eq:Ecore}
\end{equation}
where $\alpha \approx 0.1$--$0.5$ encodes short-range, atomistic core corrections. 
The dimensionless parameter $\alpha$ accounts for the short-range, non-elastic contribution to the dislocation core energy, which is not captured by continuum elasticity. Classical dislocation theory and atomistic calculations show that the core energy is approximately linear in the dislocation length and can be expressed as a fraction of $G b^{2}$, with typical values in the range $\alpha \sim 0.1$--$0.5$, depending weakly on crystal structure and bonding~\cite{HirthLothe2017,Seeger1961,Kleinert1989,BulatovCai2006}.

For a circular loop,
\begin{equation}
E_{{\rm loop},{\rm tot}}(R)
= 2\pi R \left[
\frac{G b^{2}}{2(1-\nu)} \ln\!\left(\frac{R}{r_c}\right)
+ \alpha\,G b^{2}
\right]
\sim R,
\label{eq:El_tot}
\end{equation}
so that larger loops cost more energy. It is convenient to write this in the compact form
\begin{equation}
E_{\rm loop}(R) \equiv E_{{\rm loop},{\rm tot}}(R)
= 2\pi R\, G b^{2} D(R),
\label{eq:E_D_def}
\end{equation}
with
\begin{equation}
D(R) =
\frac{1}{2(1-\nu)}
\ln\!\left(\frac{R}{r_c}\right)
+ \alpha.
\label{eq:D_def}
\end{equation}


It is useful to collect prefactors and write
\begin{equation}
E(R)
= G b^{2} R
\left[
C_1 \ln\!\left(\frac{R}{r_c}\right)
+ C_2
\right],
\label{eq:E_compact}
\end{equation}
with
\begin{equation}
C_1 = \frac{\pi}{1-\nu},
\qquad
C_2 = 2\pi \alpha.
\label{eq:C1C2}
\end{equation}
These are equivalent parametrizations of the same loop energy.

The free energy of a loop of radius $R$ at temperature $T$ is
\begin{equation}
F_{\rm loop}(R,T) = E_{\rm loop}(R) - T S_{\rm loop}(R),
\label{eq:Floop_def}
\end{equation}
which, with the expressions above, becomes
\begin{equation}
F(R,T) = E(R) - T S(R),
\label{eq:F}
\end{equation}
where $E(R)$ and $S(R)$ are given by Eqs.~\eqref{eq:E_compact} and~\eqref{eq:S}. 

Three-dimensional melting is identified with the condition that loops of arbitrarily large radius can proliferate. This corresponds to the threshold condition
\begin{equation}
F_{\rm loop}(R,T_m) = 0.
\label{eq:Fzero}
\end{equation}

We now introduce the central dimensionless quantity,
\begin{equation}
\mathcal{E}(R) \equiv \frac{E(R)}{k_B T_m},
\label{eq:Edim_def}
\end{equation}
i.e., the ratio of loop energy to thermal energy at the melting temperature. Using Eq.~\eqref{eq:ETS}, we immediately find
\begin{equation}
\mathcal{E}(R)
= \frac{E(R)}{k_B T_m}
= \frac{T_m S(R)}{k_B T_m}
= \frac{S(R)}{k_B}.
\label{eq:Edim_simple}
\end{equation}
Thus at the melting threshold, where $F(R,T_m)=0$, the dimensionless energy is simply the entropy in units of $k_B$.

Using Eq.~\eqref{eq:S} for $S(R)$, we obtain the explicit expression
\begin{equation}
\mathcal{E}(R)
= \frac{S(R)}{k_B}
= \frac{2\pi R}{a} \ln z.
\label{eq:Edim_R}
\end{equation}
Equivalently, using Eqs.~\eqref{eq:E_D_def} and~\eqref{eq:kBT_D} at the threshold $F_{\rm loop}=0$, we can write
\begin{equation}
\frac{E_{\rm loop}(R)}{k_B T_m}
= \frac{2\pi R\,G b^{2} D(R)}{a\,G b^{2} D(R)/\ln z}
= \frac{2\pi R}{a}\,\ln z.
\label{eq:Edim_ratio}
\end{equation}
This is precisely the central result of this paper:
\begin{equation}
\frac{E_{\rm loop}}{k_B T_m}
= \frac{2\pi R}{a}\,\ln z,
\label{eq:El_over_kBTm_pdf}
\end{equation}
\emph{the shear modulus $G$, Poisson ratio $\nu$, and other chemistry-dependent details cancel out entirely}, leaving a purely geometric quantity determined by $R/a$ and the local coordination encoded in $z$.

Using Eqs.~\eqref{eq:E_D_def}, \eqref{eq:D_def}, and~\eqref{eq:S}, the condition $F_{\rm loop}=0$ is
\begin{equation}
2\pi R\,G b^{2} D(R)
= T_m \cdot
\frac{2\pi R}{a}
k_B \ln z.
\label{eq:E_equal_TS_explicit}
\end{equation}
In other words,
\begin{equation}
E(R) = T_m\, S(R).
\label{eq:ETS}
\end{equation}
Solving Eq.~\eqref{eq:E_equal_TS_explicit} for $T_m$ gives
\begin{equation}
k_B T_m = a\,G b^{2}\,\frac{D(R)}{\ln z},
\label{eq:kBT_D}
\end{equation}
or explicitly
\begin{equation}
k_B T_m
= a G b^{2}
\frac{
\displaystyle \frac{1}{2(1-\nu)}
\ln\!\left(\frac{R}{r_c}\right) + \alpha
}{
\ln z
}.
\label{eq:kBTm_explicit}
\end{equation}
Solving Eq.~\eqref{eq:ETS} in terms of the compact form~\eqref{eq:E_compact} gives
\begin{align}
T_m(R)
&= \frac{E(R)}{S(R)}
= \frac{
G b^{2} R
\left[
C_1 \ln\!\left(\frac{R}{r_c}\right)
+ C_2
\right]
}{
k_B \dfrac{2\pi R}{a} \ln z
}\\
&= \frac{
G b^{2} a
}{
2\pi k_B \ln z
}
\left[
C_1 \ln\!\left(\frac{R}{r_c}\right)
+ C_2
\right].
\label{eq:Tm_general}
\end{align}
Equations~\eqref{eq:kBT_D} and~\eqref{eq:Tm_general} are equivalent representations of the melting temperature expressed in terms of elastic parameters ($G$, $\nu$ via $C_1$), core structure ($\alpha$ via $C_2$), and microscopic geometrical parameters $b$, $a$, $R$, and $z$.

\subsubsection{Configurational entropy derivation}
The configurational entropy of a dislocation loop of linear size $R$ can be estimated initially as
\begin{equation}
S_{\rm loop}(R) \sim k_B \ln\!\left(c\,\frac{R}{a} + \dots\right),
\label{eq:Slog_crude}
\end{equation}
where $a$ is a microscopic length of order the lattice spacing and $c$ is a numerical factor. A more accurate description follows from the random-walk picture of dislocation lines \cite{Gomez,HirthLothe2017}. We discretize the loop into independent segments of length $a$. The number of segments in a loop of length $L$ is
\begin{equation}
N(R) = \frac{L}{a}
= \frac{2\pi R}{a}.
\label{eq:N_segments}
\end{equation}
Each segment has $z$ distinct local configurational choices (e.g., $z \sim 4$--$6$ for typical lattices; for a simple cubic lattice, $z=6$). The number of configurations of the loop is then
\begin{equation}
\Omega(R) \sim z^{N(R)}.
\label{eq:Omega}
\end{equation}
The associated entropy is
\begin{equation}
S(R) = k_B \ln \Omega(R)
= k_B\, N(R) \ln z
= k_B \frac{2\pi R}{a} \ln z.
\label{eq:S}
\end{equation}

\section{Relation between fragility, Poisson’s ratio, and KSZ parameter}

In this Appendix we derive the relation between the kinetic fragility index $m$, the Poisson ratio $\nu$, and the repulsion steepness parameter $\lambda$ introduced in the Krausser--Samwer--Zaccone (KSZ) model. This provides a direct link between elastic properties, interaction steepness, and dynamical fragility.

An empirical correlation between the fragility index $m$ and the ratio of bulk to shear modulus was established by Sokolov and Novikov~\cite{SokolovNovikov2004}, based on experimental data for a wide class of glass-forming materials. Their relation reads
\begin{equation}
m = 29 \left( \frac{K}{G} - 0.41 \right),
\label{eq:m_KG}
\end{equation}
where $K$ is the bulk modulus and $G$ is the shear modulus.

\begin{figure}
    \centering
    \includegraphics[width=0.8\linewidth]{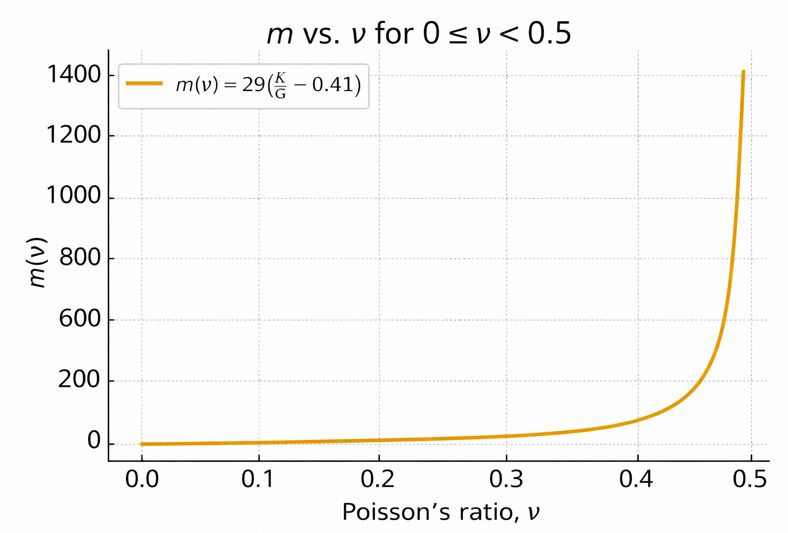}
    \caption{Fragility index as a function of Poisson ratio according to the empirical equation of Sokolov and Novikov \cite{SokolovNovikov2004}, calibrated over very many different substances.}
    \label{fig:placeholder}
\end{figure}

Within isotropic elasticity, the ratio $K/G$ can be expressed in terms of the Poisson ratio $\nu$ as
\begin{equation}
\frac{K}{G} = \frac{2(1+\nu)}{3(1-2\nu)}.
\label{eq:KG_nu}
\end{equation}
Substituting Eq.~\eqref{eq:KG_nu} into Eq.~\eqref{eq:m_KG}, we obtain the fragility explicitly as a function of $\nu$:
\begin{equation}
m(\nu) =
29 \left[
\frac{2(1+\nu)}{3(1-2\nu)} - 0.41
\right].
\label{eq:m_nu}
\end{equation}

Equation~\eqref{eq:m_nu} predicts a monotonic increase of the fragility with increasing Poisson ratio. As $\nu \rightarrow 0.5$, corresponding to the incompressible limit, the ratio $K/G$ diverges and consequently the fragility index $m$ diverges as well. This behavior reflects the increasing dominance of repulsive interactions in systems with large $\nu$.

\subsection{Relation between the KSZ parameter $\lambda$ and Poisson’s ratio}

Within the KSZ model of glassy dynamics~\cite{KSZ}, the fragility index $m$ can be expressed as
\begin{equation}
m = \frac{k_B T_g}{C_g} \, \lambda \, V_c,
\label{eq:m_KSZ}
\end{equation}
where $k_B$ is the Boltzmann constant, $T_g$ is the glass transition temperature, $V_c$ is the cooperative volume associated with shear-transformation zones, $C_g$ is a model constant, and $\lambda$ characterizes the steepness of the effective interatomic repulsion.

Equating the expressions for $m$ from Eqs.~\eqref{eq:m_nu} and~\eqref{eq:m_KSZ}, and solving for $\lambda$, we obtain
\begin{equation}
\lambda(\nu)
=
\frac{29\, C_g}{k_B T_g V_c}
\left[
\frac{2(1+\nu)}{3(1-2\nu)} - 0.41
\right].
\label{eq:lambda_nu}
\end{equation}

Equation~\eqref{eq:lambda_nu} establishes a direct relation between the KSZ repulsion steepness parameter $\lambda$ and the Poisson ratio $\nu$. The parameter $\lambda$ is a monotonically increasing function of $\nu$ and diverges as $\nu \rightarrow 0.5$, mirroring the divergence of the fragility index $m$.

\subsection{Physical interpretation}

The divergence of both $m$ and $\lambda$ in the incompressible limit reflects the underlying physical picture of the KSZ model: materials with steeper, harder interparticle repulsions exhibit higher fragility. In the extreme case of completely incompressible systems, corresponding to hard-wall interactions, the repulsion steepness becomes infinite. This limit is realized, for example, in hard-sphere amorphous packings at the jamming (random close packing) point \cite{Zaccone2022}, where the bulk modulus diverges and the Poisson ratio approaches $\nu = 0.5$.

Thus, the relations derived above provide a consistent link between elastic properties, interaction steepness, and dynamical fragility, reinforcing the interpretation of fragility as a manifestation of the microscopic steepness of interatomic repulsion.

\end{document}